\documentclass[%
 rsi,
 amsmath,amssymb,
 reprint,%
]{revtex4-2}

\usepackage{graphicx}
\usepackage{dcolumn}
\usepackage{bm}

\usepackage[utf8]{inputenc}
\usepackage[T1]{fontenc}
\usepackage{mathptmx}
\usepackage{etoolbox}
\usepackage[english]{babel}
\usepackage{siunitx}
\usepackage{amsmath,amssymb}

\makeatletter
\def\@email#1#2{%
 \endgroup
 \patchcmd{\titleblock@produce}
  {\frontmatter@RRAPformat}
  {\frontmatter@RRAPformat{\produce@RRAP{*#1\href{mailto:#2}{#2}}}\frontmatter@RRAPformat}
  {}{}
}%
\makeatother
\begin{document}

\preprint{AIP/123-QED}

\title
[A minimalist self-differencing gating scheme for dead-time-free single-photon avalanche diodes at high repetition rate]
{A minimalist self-differencing gating scheme for dead-time-free single-photon avalanche diodes at high repetition rate}

\author{Samuele Altilia}
\affiliation{Dipartimento di Fisica, Università degli Studi di Milano, Via Celoria 16, Milan, Italy}
\affiliation{Istituto Nazionale di Fisica Nucleare, Sezione di Milano, Via Celoria 16, Milan, Italy}
\author{Edoardo Suerra}
\email{edoardo.suerra@unimi.it}
\affiliation{Dipartimento di Fisica, Università degli Studi di Milano, Via Celoria 16, Milan, Italy}
\affiliation{Istituto Nazionale di Fisica Nucleare, Sezione di Milano, Via Celoria 16, Milan, Italy}
\author{Stefano Capra}
\affiliation{Dipartimento di Fisica, Università degli Studi di Milano, Via Celoria 16, Milan, Italy}
\affiliation{Istituto Nazionale di Fisica Nucleare, Sezione di Milano, Via Celoria 16, Milan, Italy}
\author{Giacomo Secci}
\affiliation{Dipartimento di Fisica, Università degli Studi di Milano, Via Celoria 16, Milan, Italy}
\affiliation{Istituto Nazionale di Fisica Nucleare, Sezione di Milano, Via Celoria 16, Milan, Italy}
\author{Sebastiano Corli}
\affiliation{Dipartimento di Fisica, Università degli Studi di Milano, Via Celoria 16, Milan, Italy}
\affiliation{Istituto Nazionale di Fisica Nucleare, Sezione di Milano, Via Celoria 16, Milan, Italy}
\author{Stefano Olivares}
\affiliation{Dipartimento di Fisica, Università degli Studi di Milano, Via Celoria 16, Milan, Italy}
\affiliation{Istituto Nazionale di Fisica Nucleare, Sezione di Milano, Via Celoria 16, Milan, Italy}
\author{Alessandro Ferraro}
\affiliation{Dipartimento di Fisica, Università degli Studi di Milano, Via Celoria 16, Milan, Italy}
\affiliation{Istituto Nazionale di Fisica Nucleare, Sezione di Milano, Via Celoria 16, Milan, Italy}
\author{Enrico Prati}
\affiliation{Dipartimento di Fisica, Università degli Studi di Milano, Via Celoria 16, Milan, Italy}
\affiliation{Istituto Nazionale di Fisica Nucleare, Sezione di Milano, Via Celoria 16, Milan, Italy}
\author{Simone Cialdi}
\affiliation{Dipartimento di Fisica, Università degli Studi di Milano, Via Celoria 16, Milan, Italy}
\affiliation{Istituto Nazionale di Fisica Nucleare, Sezione di Milano, Via Celoria 16, Milan, Italy}

\date{\today}

\begin{abstract}
Gated quenched SPAD detectors are widely used in quantum communication and quantum computing setups employing high-repetition-rate lasers. 
Here, we present a novel scheme for high-repetition-rate (\SI{100}{\mega\hertz}) sine-wave gated SPADs, based on the self-differencing technique, which significantly simplifies previous designs while offering additional advantages.
These include straightforward implementation, more precise control of the SPAD biasing, and an improved SNR. 
We implemented this approach using an InGaAs photodiode and characterized it experimentally with \SI{100}{\mega\hertz} attenuated laser pulses, measuring quantum efficiency, dark count rate, and afterpulsing behavior. 
Importantly, we demonstrate that the detector recovers full quantum efficiency in less than one pulse-repetition period after a detection event, enabling continuous operation at \SI{100}{\mega\hertz}, which, in principle, could reach the \SI{}{\giga\hertz} regime.
\end{abstract}

\maketitle

\section{Introduction}

In the era of quantum technologies, the ability to detect single photons \cite{Eisaman2011} has become a crucial requirement for a wide range of applications. These include advanced microscopy \cite{Helmchen2005}, biomedical imaging and diagnostics \cite{Lozovoy2023}, metrology \cite{Giovannetti2004}, and quantum technologies. 
In the latter domain, single-photon detection is an enabling resource both for quantum communication \cite{Gisin2002,Hadfield2009,Cavaliere2020} and for photonic quantum computing, where it underpins the realization of scalable and fault-tolerant architectures \cite{Obrien2009,Blair2025}.

A widely adopted approach for single-photon detection relies on avalanche photodiodes operated in the so called Geiger mode (SPADs, Single-Photon Avalanche Diodes) \cite{Itzler2007,Itzler2011,Jiang2007}. In this mode, the SPAD is biased above its breakdown voltage, so that the absorption of a single photon can trigger an avalanche of charge carriers, resulting in a macroscopic detectable signal. The avalanche must then be quenched by a circuit, either passive or active, that lowers the bias voltage below breakdown and subsequently restores it to its original value, recharging the diode junction capacitance \cite{Cova1996,Tosi2009,Gallivanoni2010}.

Due to their relative ease of fabrication and low cost, SPADs are among the most commonly used technologies in practical applications. Nevertheless, they are subject to some intrinsic limitations and technical challenges. First, they are affected by dark counts, spurious detection events that occur even in the absence of incident photons, primarily caused by thermally generated carriers or tunneling effects. Additionally, imperfections in the semiconductor structure can trap carriers during an avalanche and release them at a later time, producing spurious signals known as afterpulses. The probability of afterpulsing increases with both the duration and the intensity of the avalanche, and must be carefully managed to avoid degradation of detection performance. Finally, SPADs exhibit a finite dead time after each detection event, during which the device is insensitive to incoming photons. This interval, determined by the quenching and recharge processes, directly limits the maximum achievable count rate under high photon flux.

For many applications, a detector capable of detecting periodic light pulses at high repetition rates is required. The most suitable technique in this context is sine-wave gating, in which the detector is biased above the breakdown voltage by means of a periodic oscillating signal, which peaks around the expected arrival time of the pulses. Quenching is thus automatically achieved by limiting the avalanche duration inside the peak of the gating signal, and there is no need for a feedback circuit to rapidly recharge the junction. This approach enables very short recovery times with a simple design. The probability of afterpulsing can also be effectively controlled by tuning the gate peak duration, and the dark count rate is reduced compared to free-running operation, as the detector is sensitive to photons only during a small portion of the total time.

The main drawback of this solution is that, due to the parasitic capacitance of the SPAD, part of the gating signal inevitably leaks to the output, overwhelming the avalanche signals.
Several approaches have been proposed to suppress this spurious component, the most common at high gating frequencies being the self-differencing method \cite{Yuan2007,Zhang2015}.
In this technique, the output signal is split into two paths and then recombined after introducing a delay in one branch, so that the periodic gate signal cancels out through interference with its own copy.
However, practical implementation of this method is not straightforward and demands careful attention, especially to ensure precise balancing and timing between the two paths.

In this work, we present a simplified gated-quenching scheme based on a coaxial-cable delay line, which provides straightforward implementation, precise bias control, and improved SNR. 
We experimentally characterize an InGaAs SPAD operated at 100 MHz, demonstrating dead-time-free operation. This minimalist architecture offers a practical alternative to more complex self-differencing circuits, with advantages for scalable quantum communication and computing setups. 
While here we focus on a \SI{100}{\mega\hertz} implementation, the proposed approach is intrinsically scalable and, in principle, can be extended towards the \SI{}{\giga\hertz} regime, 
thus addressing the demands of high-throughput applications in quantum communication, photonic quantum computing, and biomedical imaging.

The detector design is described in Section \ref{sec:description_of_the_detector}, and its experimental characterization is presented in Section \ref{sec:characterization_of_the_detector}.

\section{Description of the detector}
\label{sec:description_of_the_detector}

\begin{figure}
    \centering
    \includegraphics[width=0.4\textwidth]{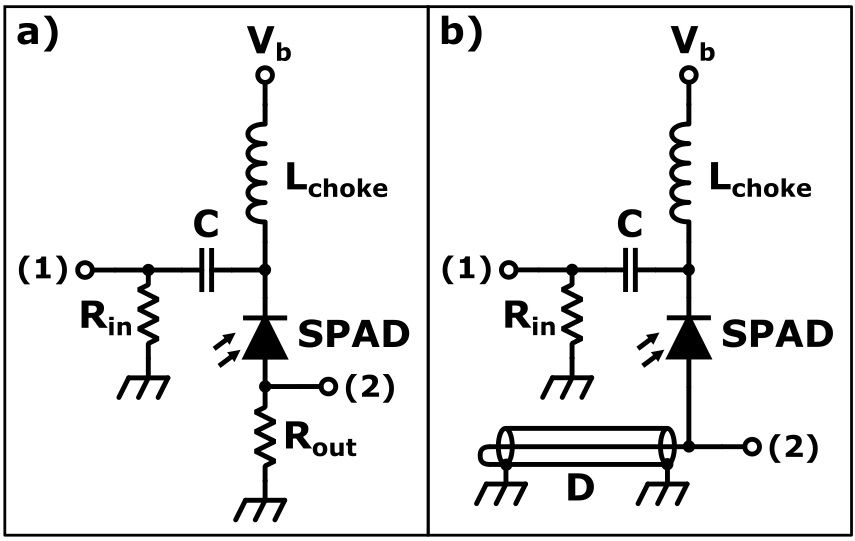}
    \caption{a) Generic scheme of a detector based on sine-wave gating. 
    b) Our solution designed to suppress the residual RF component.}
\label{fig:detector_scheme}
\end{figure}

The basic schematic of a detector based on sine-wave gating is shown in Figure~\ref{fig:detector_scheme}a). The photodiode is reverse-biased with a voltage V\textsubscript{b}, to which a periodic RF signal (typically sinusoidal) is added, reaching its peak in correspondence with the laser pulses. The RF signal, usually few volts peak-to-peak, is applied at (1), AC-coupled via the capacitor C, while the resistor R\textsubscript{in} provides impedance matching at the input. The choke inductor L\textsubscript{choke} prevents the RF signal from propagating back into the DC bias V\textsubscript{b}. Around the RF peaks, the photodiode is biased beyond its breakdown voltage, and any avalanche event produces a voltage spike across R\textsubscript{out} and so at (2). 
The main issue with this scheme is that, due to the junction capacitance of the diode (usually of the order of \SI{1}{\pico\farad}), a copy of the RF signal also appears at (2), leading the gate signal by \SI{90}{\degree}, and typically much larger in amplitude than the actual signal to be detected. This unwanted RF must therefore be canceled so that the signal can be passed to a subsequent stage, to generate a logic pulse for each avalanche. Several solutions have been proposed for this purpose, already summarized in \cite{Zhang2015}, where the output signal is processed with more or less complex circuits.

The scheme we propose, shown in Figure~\ref{fig:detector_scheme}b), represents a simplification of the self-differencing method, in which the output signal is made interfere with a suitably delayed copy of itself, so that the unwanted RF is suppressed.
To do that, two approaches are possible: summing the signal with a copy delayed by half a period and with the same sign; summing the signal with a copy delayed by a full period, but with inverted sign.
We realized that an effective way to achieve this is by simply replacing the output resistor with a properly tuned delay line (D), implemented using a simple coaxial cable impedance-matched to the output stage connected at node (2).
For the first method, it should have an open end and a length equal to one quarter of the gate period (in this case, the photocurrent must be able to discharge into the output stage connected at (2)); for the second method, it must be short-circuited at one end and have a length equal to half the gate period.
However, not only the RF signal is reflected by the delay line, but a possible avalanche peak does as well. As a consequence: in the first case two positive peaks appear in (2) for each avalanche; in the second case, a positive peak is followed by a negative one.
We decided to follow the second approach, as the resulting output signal is easier to process electronically in the following stages. This approach naturally leads to the use of a gate RF with twice the frequency of the source's repetition rate, in order to prevent dead time: it in fact avoids the risk that a delayed (negative) copy of an avalanche might cancel a subsequent one. At the same time, this brings an additional benefit, as the higher gate frequency contributes to reducing afterpulses by shortening the quenching time.

In addition to its simplicity, this scheme also offers some electronic advantages. The impedance seen from the SPAD anode to ground is ideally zero for both the DC bias and the gate RF signal, ensuring improved control over the junction voltage. Especially at high count rates, in fact, variations in the average current through the SPAD would produce voltage drops across any series resistance, degrading bias control.
Moreover, the signal extracted at point (2) can be fed directly into an amplification stage located physically very close to the SPAD anode, which can significantly improve the SNR.
Finally, if a fixed-width logic pulse is desired for each event, a fast Schmitt trigger can be used: the positive and negative peaks trigger the upper and lower thresholds respectively, generating an output pulse with a duration equal to half the source’s repetition rate.

\begin{figure}
    \centering
    \includegraphics[width=0.4\textwidth]{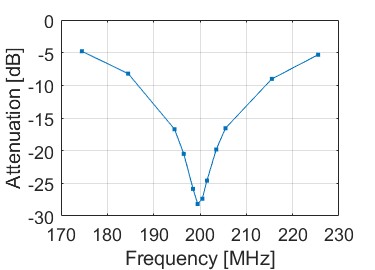}
    \caption{Attenuation of a notch filter made of a \SI{50}{\centi\meter} short-circuited RG316 cable as a function of frequency. The attenuation relative to the case without the cable is measured by connecting it in parallel to a 50-ohm input of an oscilloscope and sending a RF signal from a signal generator.}
\label{fig:attenuazione_cavo}
\end{figure}

\begin{figure}
    \centering
    \includegraphics[width=0.4\textwidth]{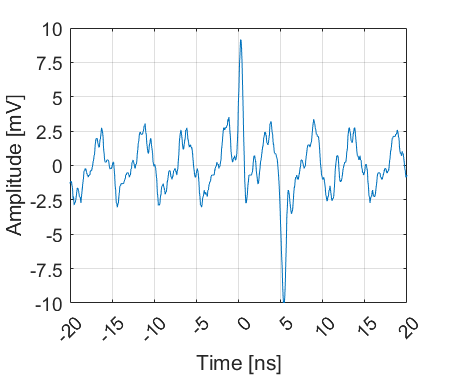}
    \caption{Example of an avalanche signal measured at the output of the assembled detector. A residual RF component can be observed, leading the gate signal peak, at which the avalanche occurs, by \SI{90}{\degree}. The reflected pulse from the delay line is also visible \SI{5}{\nano\second} later.}
\label{fig:valanga}
\end{figure}

To test the scheme in Figure~\ref{fig:detector_scheme}b), we built and tested a detector based on an InGaAs photodiode (IGA-APD-GM104-TEC). 
The cable used for the gate signal cancellation is an RG316, approximately \SI{50}{\centi\meter} long. Figure~\ref{fig:attenuazione_cavo} shows a measurement of the cable attenuation as a function of frequency, obtained by connecting the cable to the \SI{50}{\ohm} input of an oscilloscope and sending an RF signal from a signal generator. The maximum attenuation is nearly \SI{30}{\decibel}.
An example of a recorded avalanche from our prototype is instead shown in Figure~\ref{fig:valanga}: although some residual RF remains, due to cable losses and imperfect length matching of the cable, the avalanche signal is clearly distinguishable. 
The chosen component values in the prototype were: $ \text{C} = \SI{1}{\nano\farad} $, $ \text{L}_\text{choke} = \SI{10}{\micro\henry} $, and $ \text{R}_\text{in} = \SI{50}{\ohm} $, in order to match the impedance of the RF generator. The \SI{200}{\mega\hertz} RF signal was generated with a frequency doubler driven by a \SI{100}{\mega\hertz} signal extracted from a photodiode monitoring part of the laser, and is therefore intrinsically synchronized with it. Temporal matching of the RF signal with the laser pulses was then achieved using a tunable delay line on the RF signal.

\section{Characterization of the detector}
\label{sec:characterization_of_the_detector}

In this section, we characterize the detector in terms of dark count rate (DCR), quantum efficiency (QE), afterpulsing probability (APP), and dead time (DT). Before presenting the results, we first describe the operating conditions of the detector and the implementation of the key parameters, namely: the diode temperature (T), the overvoltage (OV), and the gate RF signal.
T is set and actively stabilized using a PID controller, resulting in negligible regulation error. The OV is determined with an estimated uncertainty of approximately \SI{0.1}{\volt}, due to the uncentainty in determining the breakdown voltage of the junction at the chosen T. The latter is identified by gradually increasing the bias voltage until the first avalanche events are observed in the output signal.
Finally, in our case the gate signal is generated by electronically processing the output of a photodiode that collects a fraction of the optical source, and is therefore inherently synchronized with it. The latter is a mode-locked laser operating at \SI{1030}{\nano\meter}, producing ultrashort pulses (approximately \SI{100}{\femto\second}) at a repetition rate $\nu_r = \SI{100}{\mega\hertz}$. The gate signal used throughout the characterization is a sinusoidal waveform at $2\nu_r = \SI{200}{\mega\hertz}$, with negligible higher-order harmonics and an amplitude of approximately \SI{4.0}{\volt\textsubscript{pp}}.

\subsection{Dark Count Rate}

\begin{figure}
    \centering
    \includegraphics[width=0.45\textwidth]{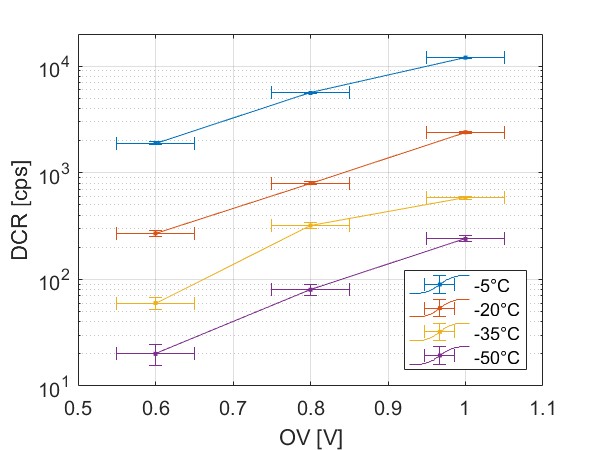}
    \caption{DCR of the detector at different OV and T.}
    \label{fig:dark}
\end{figure}

To measure the DCR of the detector we amplified its output signal and fed it into a fast comparator, followed by a digital counter. The counter recorded the number of pulses within one-second acquisition windows, from which we computed the DCR mean value and standard deviation.
Measurements were carried out for every combination of $T = \SI{-5}{\degreeCelsius}, \SI{-20}{\degreeCelsius}, \SI{-35}{\degreeCelsius}, \SI{-50}{\degreeCelsius}$ and $\text{OV} = \SI{0.6}{\volt}, \SI{0.8}{\volt}, \SI{1.0}{\volt}$. 
The results are shown in Figure~\ref{fig:dark}, which shows that the DCR increases approximately exponentially with both T and OV. A DCR of only a few hundred counts per second is already achievable at $T = \SI{-20}{\degreeCelsius}$. Furthermore note that, since the detector is active every \SI{5}{\nano\second} (corresponding to the \SI{200}{\mega\hertz} gating frequency) but is intended for use with a pulsed source at \SI{100}{\mega\hertz}, it can be reduced by a factor of two using a simple coincidence circuit with the source.

\subsection{Quantum efficiency and afterpulse probability}

We measured the QE and APP of the detector using the method introduced in~\cite{Silva2011} and further applied e.g. in~\cite{Humer2015}, which enables both quantities to be conveniently extracted from the histogram of time intervals between successive detection events. Before presenting the results, we briefly review the underlying theory, adapting it to the specific case of the pulsed regime.
In this scenario, the assumptions under which the counting statistics can be considered Poissonian (namely, that the detection time can be divided into an arbitrarily large number of time intervals with an arbitrarily small probability of detecting a pulse) are no longer valid; instead, the statistics more strictly follows a binomial distribution.

We consider a train of attenuated laser pulses indexed by $n \geq 0$, impinging on a detector that is active in correspondence with each pulse. We denote by $p(n)$ the probability of obtaining the first count in correspondence with the $n$-th bin, given that a count was registered at $n = 0$. Since in our case the DCR is generally negligible compared to the measured count rates, we assume that $p(n)$ is determined solely by the probability $p$ of having a real count in response to a laser pulse (assumed equal for every pulse), and by the probability $p_a(n)$ of having the first afterpulse in the $n$-th bin.
It can be written as
\begin{equation}
    p(n) = [1-(1-p)(1-p_a(n))] \cdot\prod_{k=1}^{n-1}(1-p)(1-p_a(k)) \,,
\end{equation}
which corresponds to the probability of having a count in the $n$-th bin multiplied by the probability of having no counts in the previous ones.
If $p \neq 0$ this is a proper probability distribution, and its normalization condition reflects the fact that every pulse must eventually be followed by another.
We can also assume that the afterpulsing probability is significant only up to a certain bin $n_a$, after which the electron traps in the junction are, with high probability, empty. Therefore, for $n > n_a$, we have $p_a(n) \simeq 0$, and we can write
\begin{equation}
    p(n)_{n > n_a} = A\,p\,(1-p)^{n-1},
\end{equation}
being $A = \prod_{n=1}^{n_a} (1-p_a(n))$.
With an exponential fit of $p(n)$ in the region where afterpulsing is negligible, it is thus possible to extract $A$ and $p$. The former can be shown (see \ref{appendix_B}) to be related to the APP through the relation $ APP = \frac{1 - A}{A}$, which can therefore be determined even without knowing the exact shape of the distribution $p_a(n)$. From the latter, instead, the QE of the detector can be inferred, provided the average photon number per pulse $\mu$ is known. Indeed, for a coherent state impinging on an on/off detector, the probability of a detection event, which corresponds to the previously defined $p$, is given by
\begin{equation}
\label{eq:detection_probability}
    p = 1 - e^{-\mu\,\text{QE}} \,,
\end{equation}
from which the QE can be obtained by inversion.

\begin{figure}
    \centering
    \includegraphics[width=0.45\textwidth]{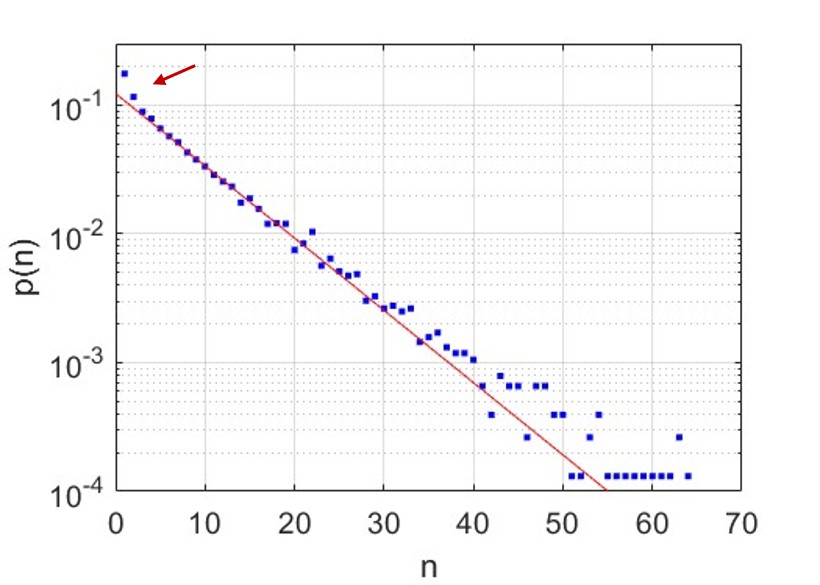}
    \caption{Measured time distribution $p(n)$ of two successive detection events, for $\text{T} = \SI{-5}{\degreeCelsius}$ and $\text{OV} = \SI{1}{\volt}$. The afterpulse region (red arrow) is clearly identifiable; beyond this region, the distribution closely follows an exponential decay (red fit).}
\label{fig:p_vs_n}
\end{figure}

To perform this characterization on our setup, the detector output signal was directly acquired by an oscilloscope in \SI{1}{\micro\second} windows (100 bins) and processed by LabVIEW to extract the arrival time statistics $p(n)$. We attenuated the laser pulses from the source using calibrated neutral density filters, to an average number of photons per pulse of $\mu = 0.59$, and sent them to the detector. We selected the order of magnitude of this number so that the arrival time distribution would decay sufficiently fast within the \SI{1}{\micro\second} window to allow proper normalization, yet much more slowly than the afterpulse distribution. This condition is necessary to enable the identification and exclusion of afterpulses from the fit, as previously described (see, e.g., Figure~\ref{fig:p_vs_n}). We found $n_a = 5$ to be a suitable choice for all fits.
Since source pulses occur every \SI{10}{\nano\second} while the detector is active every \SI{5}{\nano\second}, the time interval between the first and second detected pulses in each window was rounded to the nearest multiple of \SI{5}{\nano\second}: if this multiple was odd the event was discarded, since we are only interested in events in coincidence with the source; otherwise, it was included in the arrival time statistics. The only residual possible artifact arises when both counts are afterpulses, the first triggering the second: this leads to a slight overestimation of the APP while leaving the extracted QE unaffected. Since this effect is negligible in our case, the method was applied without the need to implement a coincidence circuit with the source.

\begin{figure}
    \centering
    \includegraphics[width=0.45\textwidth]{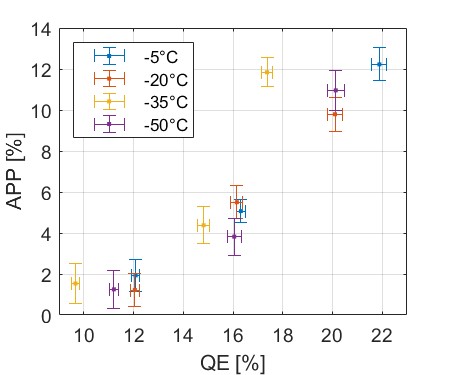}
    \caption{APP as a function of QE for different values of T. Error bars represent the standard deviations propagated from the fits used to obtain each point.}
\label{fig:APP_vs_QE}
\end{figure}

\begin{figure}
    \centering
    \includegraphics[width=0.4\textwidth]{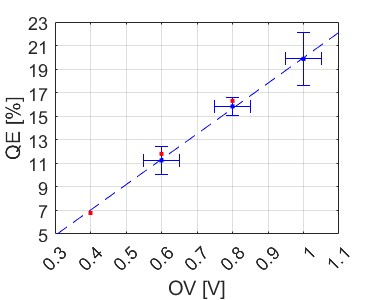}
    \caption{QE as a function of OV, obtained by averaging the results at different T. The error bars on the QE represent the semi-dispersion of the averaged values. A linear fit is shown as a dashed line. For the red dots see Figure~\ref{fig:counts_vs_photons}.}
\label{fig:QE_vs_OV}
\end{figure}

As for the DCR, we collected data for every combination of $ T = \SI{-5}{\degreeCelsius}, \SI{-20}{\degreeCelsius}, \SI{-35}{\degreeCelsius}, \SI{-50}{\degreeCelsius} $ and $ \text{OV} = \SI{0.6}{\volt}, \SI{0.8}{\volt}, \SI{1.0}{\volt} $. As a first result, we report in Figure~\ref{fig:APP_vs_QE} the trend of the APP as a function of the QE. Three distinct regions of points can clearly be identified in the graph, corresponding to the three different OV values, while the correlation with T does not appear to be significant. Therefore, we also plot in Figure~\ref{fig:QE_vs_OV} the trend of the QE as a function of the OV, averaging over the four measurements taken at different T for each OV. The points are well fitted by the line $\text{QE\,[\%]} = 19.7 \,\, \text{OV\,[V]} - 0.81$, where the nonzero intercept is evidently a trace of the error in the determination of the OV previously discussed.

\subsection{Dead time}

An additional test we report concerns the performance of the detector under extremely high repetition rate conditions. We demonstrate here that the detector behaves as an ideal one with respect to DT, meaning that it is able to fully recover its QE after a detection within a time shorter than the interval between two subsequent source pulses.

For a train of pulses impinging on an on/off detector with no DT, the probability of a detection is given by Eq.~\ref{eq:detection_probability}. Therefore, the expected behavior of the count rate $N_c$ as a function of the photon rate $N_\text{ph}=\mu \nu_r$ is \cite{Castelletto2000}:
\begin{equation}
\label{eq:counts_vs_photons}
    N_c = \nu_r \left( 1 - e^{-N_{\text{ph}} \,\text{QE}/\nu_r} \right) \,,
\end{equation}
where $\nu_r$ is again the repetition rate of the source.

If, instead, the detector becomes inactive for $n_d$ pulses after each detection, one can easily find that Eq.~\ref{eq:counts_vs_photons} must be modified as:
\begin{equation}
\label{eq:counts_vs_photons_dead_time}
    N_c^{\text{DT}} = \frac{N_c}{1+N_c n_d / \nu_r} \,.
\end{equation}
\begin{figure}
    \centering
    \includegraphics[width=0.45\textwidth]{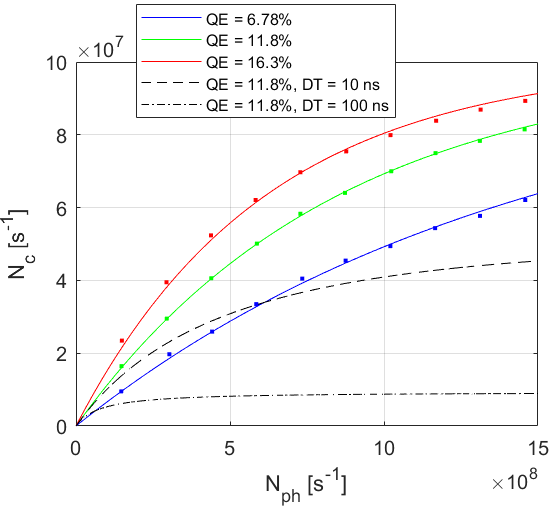}
    \caption{Detector counts per second as a function of the average number of incident photons per second, fitted using Eq.~\ref{eq:counts_vs_photons}. The three differently colored curves correspond to $\text{OV} = \SI{0.4}{\volt}, \SI{0.6}{\volt}, \SI{0.8}{\volt}$, respectively. The extracted QEs are consistent with the line in Fig.~\ref{fig:QE_vs_OV}, where they are represented as red dots. The dashed curves illustrate how the data are incompatible with even a single-pulse DT.}
\label{fig:counts_vs_photons}
\end{figure}
Figure~\ref{fig:counts_vs_photons} shows the measured count rates as a function of the of incident photons rate for $\text{OV} = \SI{0.4}{\volt}, \SI{0.6}{\volt}, \SI{0.8}{\volt}$. The photon flux was adjusted by sending the laser through a half-wave plate and a PBS, and then attenuated using calibrated neutral density filters. Measurements were performed at $\text{T} = \SI{-20}{\degreeCelsius}$, which is sufficient to ensure a negligible DCR compared to the measured rates. The experimental curves are fitted using Eq.~\ref{eq:counts_vs_photons}, showing excellent agreement. The addition of a DT effect in the model (dashed lines), as described by Eq.~\ref{eq:counts_vs_photons_dead_time}, is completely incompatible with the experimental data. This demonstrates that the detector recovers its quantum efficiency by the next pulse after each detection, and can therefore be treated as an ideal detector in pulsed operation in terms of DT. Finally, we note that the QEs obtained at different OV values are consistent with the plot in Fig.~\ref{fig:QE_vs_OV}, where they are shown as red dots.

\section{Conclusions}

We have designed and built a novel scheme for a gated quenched SPAD detector, representing a simplification of commonly used self-differencing techniques, and characterized it using a \SI{100}{\mega\hertz} repetition rate source. 
The device exhibits good performance in terms of quantum efficiency, dark count rate, and afterpulsing. 
Moreover, we demonstrate its recovery time to be below the pulse period, enabling continuous operation at \SI{100}{\mega\hertz}. 
These features make the proposed design a promising solution for high-speed quantum technologies, offering a simple and integrable approach to SPAD operation at high repetition rates.

\begin{acknowledgments}
This work has been supported by INFN group V within the projects T4QC and Adamant.
\end{acknowledgments}

\section*{Author Declarations}
\textbf{Conflict of Interest:} The authors declare that two patent applications covering the detector described in this work have been filed and accepted, with identification codes 102025000013111 and 102025000013093.

\section*{Data Availability Statement}
The data that support the findings of this study are available from the corresponding author upon reasonable request.

\appendix

\section{}
\label{appendix_A}

Let us show that $p(n)$ is a probability distribution if the physical conditions $0 < p \leq 1$ and $0 \leq p_a(n) \leq 1$ are satisfied. We rewrite
\begin{equation}
    p(n) = (1 - Q_n) \prod_{k=1}^{n-1} Q_k
\end{equation}
with $Q_n = (1 - p)(1 - p_a(n))$.

Since $0 \leq Q_n < 1$, it is clear that $p(n) \geq 0$ always holds. Furthermore, we have:
\begin{equation}
\begin{split}
    \sum_{n=1}^N p(n) &= (1 - Q_1) + (1 - Q_2)Q_1 + \\ & + \dots + (1 - Q_N)Q_1 \dots Q_{N-1} = \\ &= 1 - \prod_{n=1}^N Q_n
\end{split}
\end{equation}
therefore,
\begin{equation}
\begin{split}
    &\lim_{N \to \infty} \left| 1 - \sum_{n=1}^N p(n) \right| = \\ &= \lim_{N \to \infty} \left| \prod_{n=1}^N Q_n \right| \leq \lim_{N \to \infty} (1 - p)^N = 0
\end{split}
\end{equation}
which gives the normalization condition for $p(n)$.

\section{}
\label{appendix_B}

We evaluate here the total afterpulsing probability (APP) starting from the distribution $p_a(n)$.  
Suppose a primary count occurs, and no other sources of counts are present apart from afterpulsing. Then let the system evolve until a first afterpulse may occur. The probability $p_1$ of this event is given by the sum of the probabilities of having an afterpulse in the first bin, or none in the first and one in the second, and so on:  
\begin{equation}
    p_1 = p_a(1) + (1 - p_a(1))p_a(2) + \dots
\end{equation}
Once a first afterpulse has occurred, by the same argument the probability of having a second one is again $p_1$. Therefore, the probability of having 2 afterpulses is $p_1^2$, and similarly for 3, 4, etc. The APP is defined as the sum of the probabilities of all these possible events, quantifying the likelihood of observing one or more afterpulses from a single primary detection:  
\begin{equation}
    APP = p_1 + p_1^2 + p_1^3 + \dots = \frac{p_1}{1 - p_1} \,.
\end{equation}
Moreover, the expression for $p_1$, truncated at the first term containing $p_a(n_a)$, can be rewritten as
\begin{equation}
    p_1 = 1 - \prod_{n=1}^{n_a} (1 - p_a(n)) = 1 - A \,,
\end{equation}
as can be shown by induction on $n_a$.  
Therefore, the APP takes the form
\begin{equation}
    APP = \frac{1 - A}{A} \,.
\end{equation}
\\

\nocite{*}
\bibliography{bibliography}

\end{document}